# A plug-and-play framework for curvilinear structure segmentation based on a learned reconnecting regularization

Sophie Carneiro-Esteves[a,b,*], Antoine Vacavant[b], Odyssée Merveille[a]

[a]*Univ Lyon, INSA-Lyon, Université Claude Bernard Lyon 1, UJM-Saint Etienne, CNRS, Inserm, CREATIS UMR 5220, U1294,F-69100, Lyon, France*
[b]*Université Clermont Auvergne, CNRS, SIGMA Clermont, Institut Pascal, F-63000, Clermont-Ferrand, France*

## Abstract

Curvilinear structures are present in various fields in image processing such as blood vessels in medical imaging or roads in remote sensing. Their detection is crucial for many applications. In this article, we propose an unsupervised plug-and-play framework for the segmentation of curvilinear structures that focuses on the preservation of their connectivity. This framework includes an algorithm for generating realistic pairs of connected/disconnected curvilinear structures and a reconnecting regularization operator that can be learned from a synthetic dataset. Once learned, this regularization operator can be plugged into a variational segmentation scheme and used to segment curvilinear structure images without requiring annotations. We demonstrate the interest of our approach on the segmentation of vascular images both in 2D and 3D and compare its results with classic unsupervised and deep learning-based approach. Comparative evaluations against unsupervised classic and deep learning-based methods highlight the superior performance of our approach, showcasing remarkable improvements in preserving the connectivity of curvilinear structures (approximately 90% in 2D and 70% in 3D).We finally showcase the good generalizability behavior of our approach on two different applications : road cracks and porcine corneal cells segmentations.

*Keywords:* segmentation, curvilinear structures, plug-and-play, connectivity.

## 1. Introduction

Curvilinear structures are complex objects that exhibit a thin, curved, and elongated shape. They can be observed in many image processing fields, such as remote sensing (rivers and roads networks) or medical imaging (blood vessels and neurons). The detection (*i.e.*, segmentation) of curvilinear structures in images is the first crucial task of many applications such as blood flow simulation, urban planning, or autonomous navigation. Despite more than thirty years of research, it remains an open problem because of the complexity of its geometry and appearance in images. Curvilinear structures are sparsely distributed in images, they usually exhibit a high level of tortuosity, and are often organized in networks. Additionally, curvilinear structures are most of the time low-contrasted to the point that they can easily be altered by noise. Consequently, generic segmentation methods lead to poor results when applied to curvilinear structures. In particular, thin curvilinear structures are often missed, and the connectivity of curvilinear structure networks is rarely preserved. Some segmentation approaches tackling curvilinear structure specificities have been developed over the years. Regularization terms for variational formulations were proposed to capture specific geometric properties. The Mumford and Shah functional [1] was used to favor piecewise smoothness along the curvilinear structure intensity profile and promote connectivity [2]. However, solving this functional poses computational challenges, necessitating approximations [3] and reducing its interest. Merveille *et al.* proposed a directional total variation [4] aiming at denoising the curvilinear structures specifically along their main direction in order to reduce disconnections coming from isotropic denoising. These strategies employ indirect manners to ensure connectivity, given that connectivity is a complex geometric property that is challenging to model explicitly.

Recently, deep learning has shown promise in detecting curvilinear structures. Deep learning models can impose properties relevant to curvilinear structures,

---

*Corresponding author
Email address:* `sophie.carneiro@creatis.insa-lyon.fr` (Sophie Carneiro-Esteves )

such as sparsity or connectivity. For example, attention mechanisms [5] or reinforcement learning with a varying-patch sampling process [6] were proposed to specifically focus on the hierarchical structure and local contextual dependencies of curvilinear structures. Shit *et al.* proposed a loss function called ClDice [7] to penalize results based on their skeleton instead of their volume in order to promote connectivity. However, these methods rely on a significant amount of annotated data, which are difficult to obtain, especially in the field of curvilinear structures. Indeed, the annotation of curvilinear structures is a laborious and time-consuming task. This holds especially true in medical imaging, where the annotation task demands the expertise of a trained professional, making time a valuable resource. To cope with this, some algorithms were proposed to generate binary curvilinear structures[8, 9, 10, 11, 12], which are later used to train supervised segmentation models [10, 13].

Unrolling and "plug-and-play" approaches combine deep learning and variational methods to harness the strengths of both approaches. The unrolling approach consists of unfolding the iterative scheme of a variational approach in layers of a neural network [14, 15], and learning optimization parameters or regularization terms. On the other hand, plug-and-play methods keep the iterative scheme of a variational approach but replace the operator related to the regularization term by a regularization term learned separately by a neural network [16, 17].

By keeping an explicit model of the problem (*i.e.*, the data fidelity term), these hybrid methods yield more interpretable and robust results than pure deep learning frameworks while taking advantage of its power.

Both approaches rely on annotated datasets to train the model, although they differ in their methodology. The unrolling approach uses these data to learn the task of interest. Therefore it is necessary to have annotations for this dataset. The plug-and-play approach requires annotations only once to learn the regularization term, which can then be used to solve various tasks of interest, on another dataset, without the need for further annotations.

Given the lack of annotated data for curvilinear structures and the diversity of modalities, a plug-and-play framework is a more suitable choice.

Therefore, we propose to develop a hybrid plug-and-play approach dedicated to the segmentation of curvilinear structures. Contrary to pure deep learning approaches, we aim at designing an approach that is interpretable and robust and does not require annotations on the dataset of interest, which we will refer to as the target dataset. To this end, we propose learning a curvilinear structure reconnecting regularization term based on synthetic data, which is plugged into a variational framework to segment the target, non-annotated, dataset.

In a preliminary work [18], we showed that such a regularization yields promising results by successfully reconnecting fragments of curvilinear structures in 2D images.

In this article, we improve and go beyond this work by:

- introducing a strategy to learn a network that effectively reconnects fragmented curvilinear structures in both 2D and 3D images;

- incorporating this reconnecting network as a regularization term into a segmentation variational approach, enhancing the connectivity of the segmentation results;

- outperforming the current state-of-the-art methods in curvilinear segmentation when no annotation of the target dataset is available;

- validating the method on several datasets both in 2D and 3D from different domains, and providing a comprehensive evaluation of its performance.

The paper is organized as follows. In Section 2, we propose a brief review of curvilinear structure processing and plug-and-play methods. In Section 3, we describe our plug-and-play method and the proposed reconnecting regularization term. In Section 4, we present experiments and comparisons on 2D and 3D images. A conclusion is provided in Section 5.

## 2. Related Works

Curvilinear structure processing and analysis have been studied for many years, and a wide range of methodologies has been developed [19, 20]. In particular, with the emergence of deep learning and new computational possibilities, many deep learning-based approaches have recently been proposed, outperforming the previous state-of-the-art [21]. An extensive literature review of curvilinear structure processing is out of the scope of this article. In the following, we will focus on the two main aspects of our proposed method: the preservation of the topology of curvilinear structures and the plug-and-play approach in the context of curvilinear structure analysis.

### 2.1. Topology-preserving curvilinear structure segmentation

Some paradigms such as tracking or minimal path methods are intrinsically designed to preserve the topology of curvilinear structures. Tracking [22] is based on recursively recruiting neighbors pixels according to a criterion (*e.g.*, the pixel intensity). Similarly, minimal path methods [23] aim to find the best path, according to a cost function, between two points. In the context of curvilinear structures, this path is supposed to be the structure centerline. Both paradigms show interesting results; however, they need a human interaction to define a set of seed points, which can be tedious in the context of large and complex curvilinear structure networks.

Many deep learning approaches have been proposed recently to tackle curvilinear structure segmentation. In particular, U-Net [24, 25], the gold standard for medical image segmentation, is frequently used for vascular segmentation [26, 27, 28]. Among these methods, two main ideas have been explored to preserve the topology of curvilinear structures: either constraining the loss function of the network or modifying its architecture.

First, Betti numbers, representing the number of k-dimensional holes in a structure (*i.e.*, a segmentation in our case), have been used to propose topology-preserving losses. In 2D, Hakim *et al.* [29] defined a loss based on the Euler characteristic (a topological invariant defined as a linear combination of the Betti numbers) to reduce the number of isolated objects. Similarly, Hu *et al.*[30] proposed a loss based on persistent homology to extract the first Betti number corresponding to the number of connected components. Clough *et al.* [31] extended the approach to 3D images and all Betti numbers. Another work on 3D images [32] proposed to predict the persistent homology along with the segmentation to estimate the topology error and therefore improve the model. These losses evaluate whether the topology of the ground truth and the predicted segmentation match without taking into account if the underlying shapes overlap. To address this problem, Stucki *et al.* [33] proposed a topological loss based on the induced matching of persistence barcodes, ensuring penalization based on the topology but also the geometry of the segmentations.

On the other hand, losses based on the curvilinear structure centerlines, such as the ClDice [7], were proposed to foster connectivity and reduce the well-known volumetric bias existing between thick and thin curvilinear structures, leading to the vanishing of the latter. However, its use in 3D has not been successful because of the complexity to extract accurate centerlines. To tackle this issue, Rougé *et al.* [34] proposed a cascaded U-Net to learn the skeletonization operation in addition to the segmentation and use the ClDice based on these more accurate centerlines. Mosinska *et al.* [35] proposed to learn a VGG on the ImageNet dataset, which is then employed to extract features from both the segmentation prediction and its ground truth. They introduced a loss function that involves minimizing the distance between these features.

Architecture can also be adapted to take into account curvilinear structure features. Attention modules [36] have been added in the latent space of a U-Net based-architecture to promote the connectivity [5, 37]. Multitask architectures have also been proposed, adding one or several proxy tasks, such as centerline detection, that better take into account the topology of curvilinear structures [38]. Lin *et al.* [39] proposed a cascaded architecture of two U-Nets, the first one was employed to extract texture features, while the second one focused on topology correction with contrastive learning. Some works [40, 41] focused on adding a transformer architecture [36] into the bottleneck of a U-Net to make the model consider the overall context of the image. They claimed that it improved the detection of long-distance relationships between pixels, ultimately leading to the better preservation of curvilinear structures connectivity.

Many segmentation pipelines include a first step based on classic vesselness filtering [42] to enhance the contrast between curvilinear structures and the background. A few works investigated how to use these methods in a deep learning framework, in particular by using the directional information provided by these filters, and thereby promoting connectivity [43, 44]. Most of these methods were only developed for 2D images, and they require a large amount of annotated data, which limits their use in real applications.

Segmentations can also undergo post-processing techniques to reconnect interrupted segments. Different algorithms have been proposed based on the centerlines [45], graph-based approaches [46, 47], and contour completion processes [48]. These methods are complex to use, parameter-dependent, and were only proposed for 2D images.

### 2.2. Variational approaches for curvilinear structure segmentation

The variational approach for image restoration is defined as the minimization of two energies:

$$\hat{u} = \underset{u \in [0,1]^N}{\operatorname{argmin}} E_{\text{data}}(u, f) + \lambda E_{\text{reg}}(u), \qquad (1)$$

where $f \in \mathbb{R}^N$ is the initial image of $N$ pixels, $\hat{u} \in \mathbb{R}^N$ is the restored image, $E_{\text{data}}$ is the data fidelity, $E_{reg}$ the regularization term which promotes desirable solution properties and $\lambda \in \mathbb{R}$ is a regularization coefficient acting as a trade-off between both terms.

A classic regularization term used in image processing is the Total Variation (TV), which promotes a piecewise smooth solution [49]. Even though the term is efficient for denoising natural images, it tends to make thin curvilinear structures disappear, leading to a connectivity loss in many applications. To address this issue, a directional regularization adapted to curvilinear structures has been proposed [4] by incorporating prior knowledge on the curvilinear structure orientation. Even though this term promotes connectivity by denoising in an anisotropic manner, it does not lead to the reconnection of significantly separated curvilinear fragments. Moreover, this term requires the computation of a curvilinear structure estimator, such as the RORPO filter [50] or the Frangi vesselness [51], which can be time-consuming and/or complex to parametrize [42].

Outside the curvilinear structure scope, Heide *et al.* [52] proposed to replace or combine a classic regularization term, such as TV, with a denoising algorithm like BM3D [53]. BM3D was used as a self-similarity inducing denoising prior on the restored image. Following this work, a series of works [16, 54, 55, 56] has been proposed to replace the regularization term by a denoising neural network, which learns the noise distribution of a dataset of interest. Therefore, these learned regularization terms are more adapted to the application than classic regularization ones. These approaches were used to tackle image restoration applications such as demosaicking, deconvolution, inpainting, or deblurring. To the best of our knowledge, all these methods proposed to learn a prior only on the noise of the observed image. But no work has been proposed to learn more complex regularizations such as the connectivity of curvilinear structures.

This article presents the first regularization term aiming to add a more complex constraint on the solution of interest and a complete framework to perform segmentation.

## 3. Proposed method

In this section, we will first present a global overview of our proposed framework, then we will describe separately each step.

### 3.1. Overview

In this work, we aim at learning a regularization term promoting the connectivity of curvilinear structures, which is a property difficult to define explicitly. Subsequently, this term is incorporated into a segmentation plug-and-play approach.

To this end, we propose a three-step framework composed of:

- The generation of a disconnected binary curvilinear structure dataset.

  To learn a connectivity property, we propose to train a network using a synthetic dataset of paired images. This dataset comprises an input image containing disconnected curvilinear structures and its corresponding version with the structures connected (see Fig. 1, top left). To generate these images, we developed an algorithm simulating realistic disconnections in curvilinear structures.

- The reconnecting regularization term learning.

  From the previously generated dataset, we learn a network $G_{\text{reco}}$ (Fig. 1 top right) with a residual U-Net aiming at reconnecting curvilinear structures in an image (see Section 3.3).

- The plug-and-play segmentation.

  The learned network $G_{\text{reco}}$ is finally plugged into a segmentation variational framework, by replacing the proximity operator of the regularization term $E_{\text{reco}}$ in the segmentation iterative scheme (Fig. 1 bottom).

The code of our framework is available at https://github.com/creatis-myriad/plug-and-play-reco-regularization.

### 3.2. Dataset creation

Our objective is to develop a regularization term that promotes the connectivity of curvilinear structures. To be generic and usable in many different applications, this term should not depend on the input image modality or the type of curvilinear structure (*e.g.* road, vessels, *etc.*) but rather focus on the geometric properties of these structures.

Such a reconnecting regularization would be too complex to express explicitly, so we have chosen to learn it based on a generated dataset of connected/disconnected curvilinear structure images. We have opted for a synthetic binary dataset to represent the connectivity property, as binary images not only exhibit modality independence but also ease the generation of realistic disconnections.

From an initial image containing binary curvilinear structures, our algorithm creates random disconnections inside curvilinear structures according to several rules we have established based on our experience and observations of current automatic segmentation algorithm results.

More specifically, the following steps are performed to create an image containing disconnections:

1. Extraction of the curvilinear structure centerlines and its distance map. The distance map is defined for each pixel inside the curvilinear structures, and its value is the distance to the background.

2. Classification of each centerline pixel into $m \in \mathbb{N}$ classes based on their radius.

3. Selection of the $p$ ($p \leqslant m$) thinnest curvilinear structure classes to avoid disconnections on the biggest structures.

4. Generation of disconnections:

   - Selection of a value $i \in [0, p]$ drawn from the probability distribution $\mathrm{P}(i) = \frac{2^{p-i}}{2^p-1}$. Here, $i$ represents the radius class of the curvilinear structure which will be disconnected.

   - Random selection of a centerline pixel in the image corresponding to class $i$. This point will be the center of the disconnection.

   - Selection of the disconnection size $s$ from a normal distribution with a mean of $\frac{C}{i+1}$ with $C \in \mathbb{N}$ a constant depending on the dataset.

   - Generation of the disconnection
     The disconnection is generated by removing random pixels in a disc (or a ball) centered on the centerline pixel.

Afterwards, additional fragments are added to the image to mimic non-curvilinear structure fragments. These fragments are introduced to make the network learn to differentiate fragments aligned with curvilinear structures, which should probably be reconnected, from noise and non-curvilinear structures, that should not be reconnected. In 2D, the fragments are generated by adding random pixels inside a disc area. In 3D, these fragments are generated using the algorithm proposed by Douarre *et al.* [57] to generate more complex patterns.

### 3.3. Reconnecting regularization term learning

We use the previously generated dataset to learn the reconnecting operator $G_{\mathrm{reco}}$. Our model is designed to effectively reconnect fragments of curvilinear structures while removing the non-curvilinear fragments.

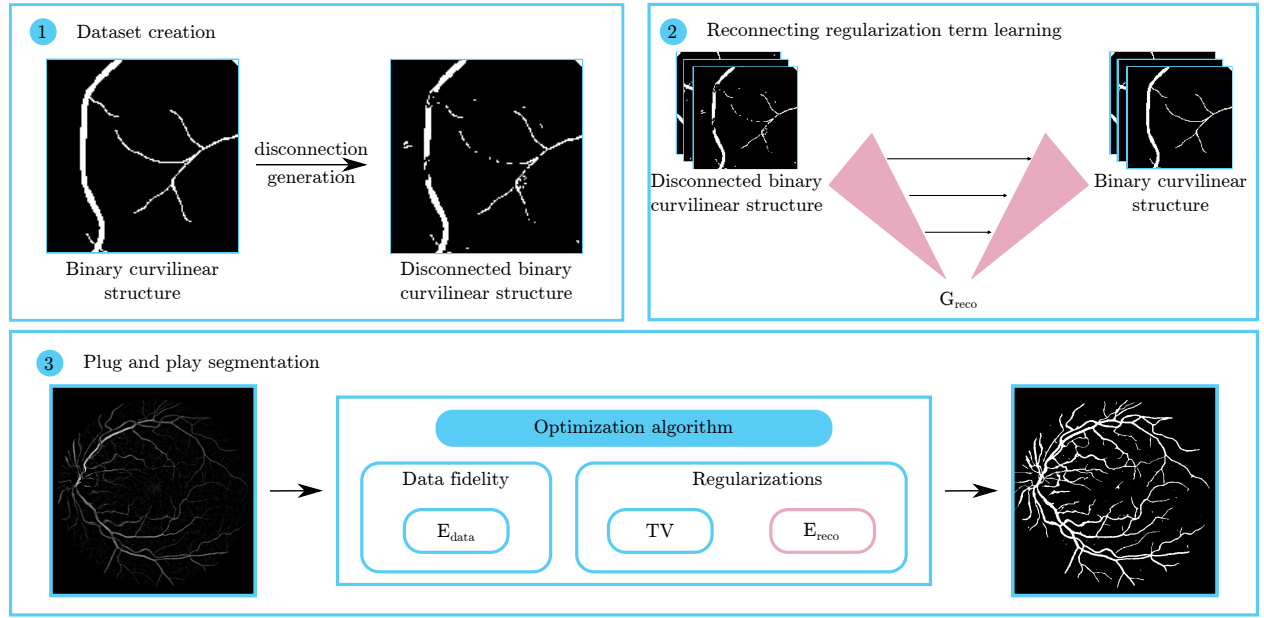

Figure 1: Pipeline of our proposed framework. First, a dataset is generated with pairs of synthetic connected/disconnected curvilinear structures. Second, this dataset is used to train a model $G_{\mathrm{reco}}$ with a residual U-Net architecture. Third, this model is finally plugged into the optimization scheme to solve the segmentation problem by replacing the proximity operator of $E_{\mathrm{reco}}$. Note that for the plug-and-play segmentation, it is essential that the input has a homogeneous background. A preprocessing step, such as subtracting the median filter of the image, may be necessary.

We trained a residual $n$D UNet ($n = 2$ or $n = 3$ depending on the image dimension) [58]. U-Net is the gold standard for medical image segmentation, and adding residual units to its architecture improves its robustness to inputs that are dissimilar to training data. This property is essential because we aim at learning a generic regularization term which will be used on different target datasets.

Our U-Net architecture is 4-layer deep using batch normalizations and 16, 32, 64 and 128 features at each layer.

In the literature, the Dice loss [59] has been widely used for learning segmentation in unbalanced classes scenarios. This is the case for our problem, where the curvilinear structures only represent a small portion of the images pixels, and the missing fragments only a small fraction of the curvilinear structures themselves.

However, only using a Dice loss would focus the network attention on preserving the curvilinear structures in general, without focusing on the accurate identification of the missing fragments.

To address this issue, we propose to use a combination of two Dice losses (see Eq. 2). The first one is a classic Dice $\mathcal{D}$ on the image to guide the reconstruction of the whole curvilinear structures. The second Dice loss, called a weighted Dice $\mathcal{D}_w$, is computed only within a mask $M$ of the missing fragments and its close neighborhood. The total loss $\mathcal{L}$ is defined as follows:

$$\mathcal{L}(x, y) = \mathcal{D}(x, y) + \mathcal{D}_w(x, y; M), \quad (2)$$

with $x$ and $y$ the image and its associated annotation.

The missing fragment mask is generated by a dilation of the missing fragments with a disc (or ball in 3D) structuring element of radius $r = 2$.

We train our models with patches of size $96^n$ voxels and performed an on-the-fly data augmentation with random rotations and flips. The learning rate is set to $10^{-3}$ and the networks is trained for 1000 epochs with an Adam optimizer. The model with the lowest validation loss is kept.

### 3.4. Plug-and-play segmentation

We express our segmentation problem as a variational approach (see Eq.1). In this section, we first present the choice of the data fidelity energy $E_{\text{data}}$ and the regularization term $E_{\text{reg}}$, then we describe the iterative optimization scheme used to solve this segmentation problem.

### 3.4.1. Data fidelity and regularization terms

We have chosen the Chan *et al.* segmentation data fidelity term [60] defined as follows:

$$\begin{aligned} E_{\text{data}}(u, f) &= \left\langle u, c_f \right\rangle_F, \\ c_f &\mapsto (c_1 - f)^2 - (c_2 - f)^2, \end{aligned} \quad (3)$$

with $c_1$ and $c_2$ constant values corresponding respectively to the mean value of the background and the foreground, and $\langle ., . \rangle_F$ is the Frobenius product.

In the previous sections, we have presented our reconnecting regularization term, $E_{\text{reco}}$ which promotes a connected segmentation. In addition to this term, we need a classic regularization/denoising term promoting a smooth segmentation. To favor such a property, we add the classic total variation, $TV(u)$, defined by:

$$TV(u) = \|\nabla u\|_{2,1}, \quad (4)$$

with $\|.\|_{2,1}$ the $l_1$ norm of the $l_2$ norm and $\nabla$ the gradient operator.

As our goal is to obtain a binary segmentation result, we need a last regularization term to constrain the values of the solution between 0 and 1. We use $\iota_{[0,1]^N}$, the indicator function of the set $[0, 1]^N$.

Our reconnecting regularization is defined based on binary images and cannot be applied directly on a greylevel image. During the iterative optimization scheme, the grey-level input image will progressively be transformed into a binary segmentation. When the image is close to a binary image, but the iterative scheme has not yet converged, we propose to introduce our learned reconnecting regularization $E_{\text{reco}}$. Our final regularization term $E_{\text{reg}}$ is defined as follows:

$$E_{reg}(u) = \begin{cases} \lambda TV(u) + E_{\text{reco}}(u) & \text{if } u \text{ almost binary,} \\ \lambda TV(u) + \iota_{[0,1]^N}(u) & \text{else.} \end{cases} \quad (5)$$

The condition *almost binary* depends on the number of iterations of the solving algorithm and is set experimentally (see Section 4).

### 3.4.2. Optimization scheme

Our segmentation problem is a three-term energy : $E_{\text{data}}$, TV and $\iota_{[0,1]^N}$ or $E_{\text{data}}$, TV and $E_{\text{reco}}$. We propose to use a *Forward-Backward Primal-Dual* (FBPD) algorithm [61] which solves the following primal problem:

$$\hat{u} = \underset{u}{\text{argmin}}\, h(u, f) + g(Lu) + k(u), \quad (6)$$

where $k$, $g$ and $h$ are lower semi-continuous convex functions from $\mathbb{R}^N$ to $(-\infty, +\infty]$. $h$ is differentiable with a $\beta$-Lipschitz continuous gradient ($\beta > 0$), and $L \in \mathbb{R}^{N^2}$ an operator and $\hat{u}$ the primal solution.

The FBPD algorithm solves the primal problem Eq.(6) by relying on its associated dual problem :

$$\hat{v} = \underset{v}{\mathrm{argmin}}(k^\star \square h^\star)(-L^T v) + g^\star(v), \quad (7)$$

where $g^\star$ the conjugate of $g$, $\square$ the inf-convolution defined as $(f \square g)(x) = \inf_{y \in \mathbb{R}^N} f(y) + g(x - y)$, and $\hat{v}$ the dual solution.

The FBPD iterative scheme is defined in Eq. 8.

$$u_{i+1} = \mathrm{prox}_{\tau k}(u_i - \tau(\nabla h(u_i) + L^T v_i)),$$
$$v_{i+1} = \mathrm{prox}_{\sigma g^\star}(v_i + \sigma L(2u_{i+1} - u_i)), \quad (8)$$

with $\mathrm{prox}_{\sigma g}$ the *proximal* operator of $\sigma g$, $g^\star$ the conjugate of $g$, $\tau \in \mathbb{R}^+$ and $\sigma \in \mathbb{R}^+$ scalar hyperparameters. $\tau$ and $\sigma$ must be set according to Eq. 9 to ensure the convergence of the algorithm.

$$\tau^{-1} - \sigma \|L\|_S^2 \geq \frac{\beta}{2}, \quad (9)$$

with $\|.\|_S$ the spectral norm.

To use the FBPD algorithm on our segmentation problem, we define the following terms:

$$
\begin{aligned}
h(u, f) &= E_{\mathrm{data}}(u) = \langle u, c_f \rangle_F, \\
g(u) &= \lambda \|u\|_{2,1}, \\
L &= \nabla, \\
k(u) &= \begin{cases} E_{\mathrm{reco}}(u) & \text{if } u \text{ almost binary,} \\ \iota_{[0,1]^N}(u) & \text{else.} \end{cases}
\end{aligned} \quad (10)
$$

To apply the FBPD algorithm, closed-form solutions of the proximity operators of Eq. 8 are required :

$$
\mathrm{prox}_{\sigma g^\star}(u) = \frac{\lambda \sigma^{-1}}{\max\left(\left\|\frac{u}{\sigma}\right\|_2, \lambda \sigma^{-1}\right)},
$$
$$
\mathrm{prox}_{\sigma \iota_{[0,1]^N}}(u) = \mathrm{proj}(u) = \begin{cases} u & \text{if } u \in [0,1] \\ 0 & \text{if } u < 0 \\ 1 & \text{otherwise} \end{cases} \quad (11)
$$

When $u$ is almost binary, following the classic plug-and-play paradigm, we replace the proximity operator $\mathrm{prox}_{\tau E_{\mathrm{reco}}}$ of the FBPD iterative scheme by our reconnecting network $G_{\mathrm{reco}}$.

Finally, our plug-and-play segmentation algorithm is presented in Algorithm 1.

---

**Algorithm 1:** Plug-and-play segmentation with the learned reconnecting operator

**Data:** $\alpha \in \mathbb{N}^{+*}$,
$\quad\quad u_0 \in \mathbb{R}^{N^2}, v_0 \in \mathbb{R}^{2N^2}, (\tau, \sigma) \in (0, +\infty)^2$
**for** $i \geq 1$ **do**
$\quad$ $p_i = (u_i - \tau(\nabla h(u_i) + L^T v_i)$
$\quad$ **if** $i < \alpha$ **then**
$\quad\quad$ $u_{i+1} = \mathrm{prox}_{\tau \iota_{[0,1]^N}}(p_i)$
$\quad$ **else**
$\quad\quad$ $u_{i+1} = G_{\mathrm{reco}}(\mathrm{proj}(p_i))$
$\quad$ $v_{i+1} = \mathrm{prox}_{\sigma g^\star}(v_i + \sigma L(2p_i - u_i))$

---

## 4. Experiments

In this section, after describing our experimental setup, we first demonstrate the interest of our reconnecting operator on 2D and 3D images of blood vessels. We compare its results with both classic and deep learning-based approaches. Then, we analyze it through an ablation study. Finally, we illustrate the interest of our approach on other applications.

### 4.1. Experimental setup

#### 4.1.1. Datasets

We chose two applications to conduct our experiments: a 2D application on retinal images, and a 3D application on injected CT-scans of the liver. One of the main interest of our approach is that it does not require annotations of the target dataset to perform the segmentation, but only synthetic images for training $G_{\mathrm{reco}}$. To illustrate this point, we needed for each application one synthetic dataset to learn our reconnecting term, and another to perform the segmentation.

In 2D, we chose to learn our reconnecting term on 20 binary curvilinear structures generated using the OpenCCO algorithm [9] (see Fig. 4(a)). Then, we generated our pairs of connected/disconnected curvilinear structure images using our algorithm to generate random disconnections (see Section 3.2) and applied random rotations, to obtain a dataset of 80 pairs of connected/disconnected curvilinear structure images. For the segmentation, we used the DRIVE dataset [62], which consists of 40 retinal images and their associated annotations.

In 3D, we trained our reconnecting term on a synthetic dataset of curvilinear structures generated with the Vascusynth software [1] [8]. We generated 315 images

---

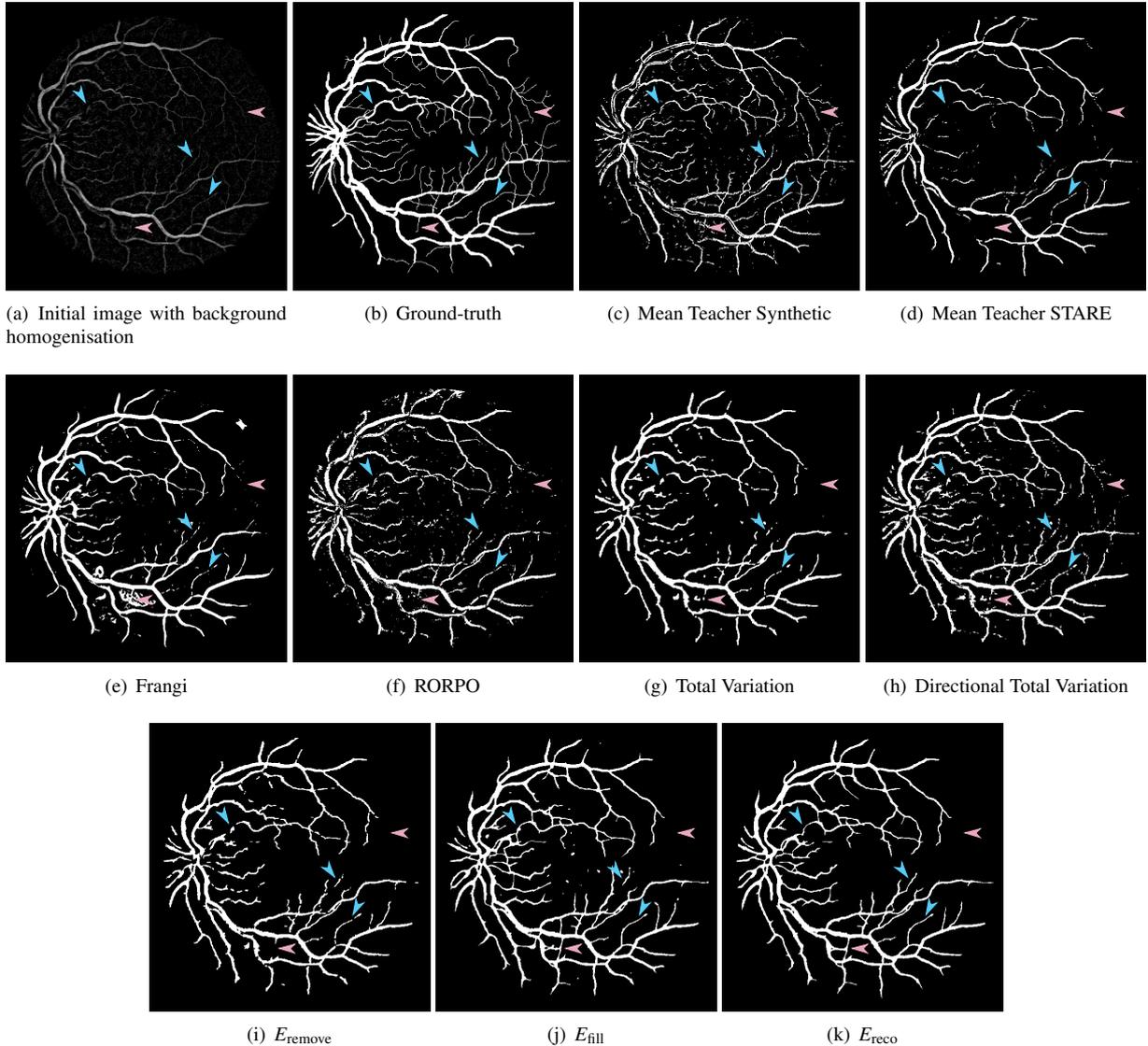

(a) Initial image with background homogenisation

(b) Ground-truth

(c) Mean Teacher Synthetic

(d) Mean Teacher STARE

(e) Frangi

(f) RORPO

(g) Total Variation

(h) Directional Total Variation

(i) $E_{\text{remove}}$

(j) $E_{\text{fill}}$

(k) $E_{\text{reco}}$

Figure 2: Comparison of segmentation results on one DRIVE image. Blue arrows point to some examples of successful reconnections of our reconnecting regularization term. Pink arrows show false reconnections or missing vessels.

that we disconnected with our proposed algorithm (see Section 3.2). As the target dataset for the segmentation, we used the IRCAD dataset[2], which is composed of 19 3D liver CT-scans and their associated annotations.

Note that the DRIVE and IRCAD annotations were only used for the evaluation of the method, and not for the training.

---

### 4.1.2. Compared methods

Firstly, we conducted a comparison between our learned reconnecting term $E_{\text{reco}}$, the classic TV regularization [49], and the directional TV [4] which is specifically designed to reconnect curvilinear structures. For a fair comparison, the segmentation with the classic TV (resp. the directional TV) is performed with a FBPD algorithm, where $g(Lu)$ is the TV (resp. the directional TV) and $k(u)$ is $\iota_{[0,1]^N}(u)$.

We set the same parameters $c_1$, $c_2$, $\sigma$ and $\tau$ for the three methods. The background and foreground constants $c_1$ and $c_2$ were set experimentally. The parame-ters $\tau$ and $\sigma$ were set respectively to 1.587 and $10^{-3}$ in 2D and to 1.493 and $10^{-3}$ in 3D based on Eq. 9.

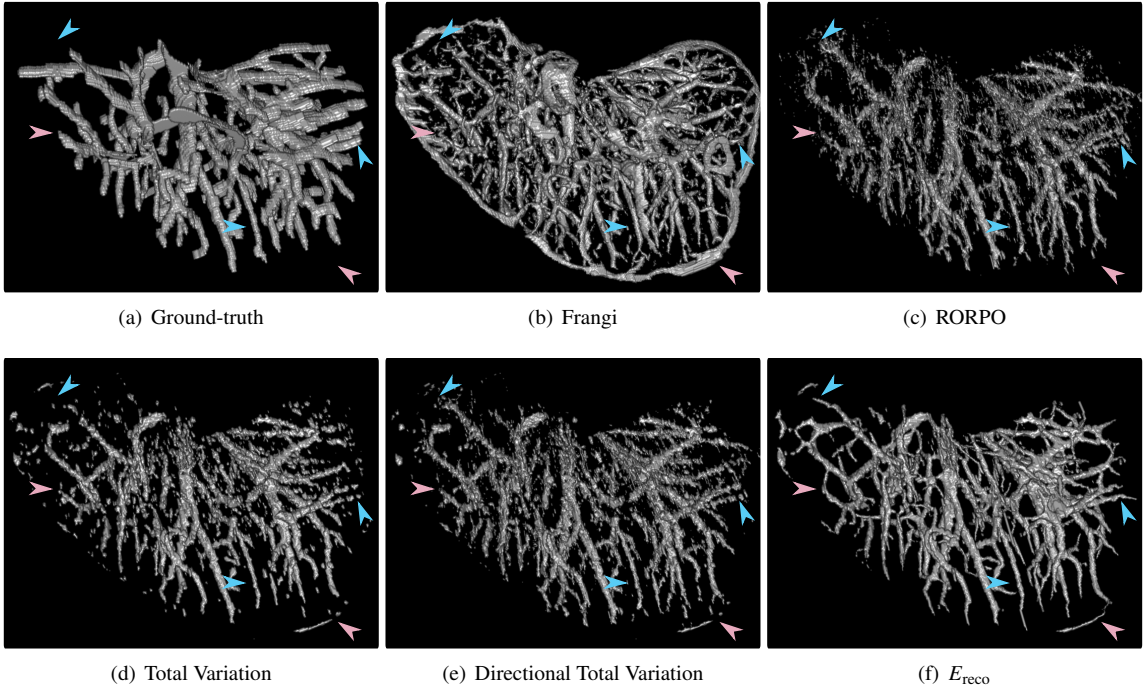

|              |              |              |
|:------------:|:------------:|:------------:|
| (a) Ground-truth | (b) Frangi | (c) RORPO |
| (d) Total Variation | (e) Directional Total Variation | (f) $E_{\text{reco}}$ |

Figure 3: Comparison of segmentation results on one IRCAD volume. Blue arrows point to some examples of successful reconnections of our reconnecting regularization term. Pink arrows show false reconnections.

To ensure a fair comparison, the values of $\lambda$ were independently optimized for each image and each method. We selected the value maximizing the MCC in the range $[0.001, 0.080]$ in 2D, and $[0.001, 0.050]$ in 3D, with a step of $10^{-3}$.

The FBPD scheme was applied until convergence or for a maximum of 1000 iterations. The switch between $u_{[0,1]^N}$ and our reconnecting regularization $E_{\text{reco}}$ in our approach is done at $\alpha = 500$ iterations when we observed that $u$ is usually almost binary. We observed that our approach is not very sensitive to this value, and it could be set approximately to half the maximum number of iteration performed.

The Chan *et al.* data fidelity term assumes that the input image has a homogeneous background. We thus preprocessed each image by subtracting its median filter. The kernel of the median filter was set once for each dataset experimentally.

Furthermore, as we developed an unsupervised approach, we compared our framework with classic unsupervised method for blood vessels filtering and segmentation : RORPO [50] and Frangi [51]. Parameter optimization was carried out using a grid search, exploring a predefined set of values for each hyperparameter to determine the optimal configuration of each method based on the MCC (see Sec 4.1.3).

Most recently proposed deep-learning-based vascular segmentation methods are primarily supervised, showcasing high performance when annotations are available. However, their performance drastically drop when annotations are lacking. Such supervised approaches are out of the scope of this article, however, for reference purposes, we provided average metrics reported in the literature, offering insights into the upper limits of performance for comparative analysis. In 2D, an average Dice of 0.836 was reported in [63], while in 3D, an average Dice of 0.838 was achieved according to [64]. In this article, we compared our results with a deep-learning based approach in similar conditions than our method, *i.e.*, in an unsupervised context, on an unlabeled target dataset. In this context, classic supervised deep learning-based segmentation strategies cannot be applied, and two scenarios may occur:

1. scenario 1: one has access to an annotated dataset similar to the target dataset.
2. scenario 2: one can generate a simulated dataset with annotation, similar to the target dataset.

We compared our results with both scenarios using a mean teacher (MT) network [65], which is a popular approach used to learn a model with both annotated and non annotated data from related datasets.

| Method | | Volumetric Metrics | | | | Geometric Metrics | | | Topological Metrics | | |
|---|---|---|---|---|---|---|---|---|---|---|---|
| | | TPR | PPV | MCC | Dice | ClDice | 95HD | ASSD | $\epsilon_{\beta_0}$ | $\epsilon_{\beta_1}$ | $\epsilon_\chi$ |
| 2D | RORPO | 0.650 ± 0.053 | 0.713 ± 0.064 | 0.636 ± 0.050 | 0.678 ± 0.044 | 0.626 ± 0.048 | 11.789 ± 3.357 | 2.293 ± 0.608 | 52.275 ± 34.062 | 0.924 ± 0.059 | 3.872 ± 1.060 |
| | Frangi | 0.716 ± 0.051 | 0.756 ± 0.042 | 0.699 ± 0.044 | 0.735 ± 0.041 | 0.717 ± 0.047 | 13.576 ± 4.441 | 2.410 ± 0.846 | 21.340 ± 15.959 | 0.646 ± 0.126 | 1.849 ± 0.794 |
| | MT synth. | 0.602 ± 0.070 | 0.864 ± 0.039 | 0.689 ± 0.047 | 0.707 ± 0.051 | 0.713 ± 0.053 | **7.268** ± 3.230 | **1.509** ± 0.511 | 22.809 ± 16.696 | 0.614 ± 0.204 | 1.906 ± 0.531 |
| | MT STARE | 0.607 ± 0.061 | **0.903** ± 0.033 | 0.711 ± 0.041 | 0.724 ± 0.044 | 0.717 ± 0.051 | 15.932 ± 3.728 | 2.421 ± 0.580 | 11.235 ± 7.725 | 0.777 ± 0.100 | 1.459 ± 0.323 |
| | TV | 0.690 ± 0.051 | 0.821 ± 0.069 | 0.719 ± 0.041 | 0.747 ± 0.036 | 0.730 ± 0.044 | 13.489 ± 5.186 | 2.265 ± 0.885 | 24.220 ± 15.885 | 0.749 ± 0.113 | 2.135 ± 0.545 |
| | Directional TV | 0.694 ± 0.053 | 0.819 ± 0.076 | 0.720 ± 0.045 | 0.748 ± 0.041 | 0.728 ± 0.049 | 12.143 ± 5.014 | 2.071 ± 0.769 | 25.833 ± 22.350 | 0.741 ± 0.103 | 2.219 ± 1.129 |
| | $E_{\text{reco}}$ | **0.725** ± 0.049 | 0.803 ± 0.069 | **0.729** ± 0.041 | **0.759** ± 0.036 | **0.744** ± 0.045 | 16.095 ± 5.235 | 2.550 ± 0.924 | **2.685** ± 2.767 | **0.316** ± 0.187 | **0.417** ± 0.232 |
| 3D | RORPO | 0.408 ± 0.121 | 0.640 ± 0.087 | 0.487 ± 0.090 | 0.488 ± 0.096 | 0.525 ± 0.104 | 13.145 ± 4.600 | 2.917 ± 0.873 | 1.029 ± 1.444 | 57.818 ± 52.836 | 66.991 ± 106.262 |
| | Frangi | **0.455** ± 0.081 | 0.492 ± 0.127 | 0.441 ± 0.087 | 0.462 ± 0.079 | 0.491 ± 0.109 | 21.939 ± 7.465 | 4.732 ± 1.491 | 5.079 ± 4.980 | 6.817 ± 16.565 | 13.284 ± 27.567 |
| | TV | 0.344 ± 0.139 | **0.721** ± 0.126 | 0.473 ± 0.116 | 0.450 ± 0.129 | 0.533 ± 0.166 | **11.623** ± 4.213 | 2.954 ± 1.017 | 2.253 ± 3.300 | **4.533** ± 10.674 | 4.789 ± 5.809 |
| | Directional TV | 0.362 ± 0.126 | 0.695 ± 0.099 | 0.477 ± 0.092 | 0.462 ± 0.105 | 0.562 ± 0.106 | 11.734 ± 4.581 | 2.666 ± 0.819 | 1.682 ± 2.266 | 41.966 ± 51.223 | 32.495 ± 42.293 |
| | $E_{\text{reco}}$ | 0.435 ± 0.153 | 0.675 ± 0.105 | **0.513** ± 0.087 | **0.507** ± 0.102 | **0.585** ± 0.079 | 13.032 ± 4.169 | 2.895 ± 0.877 | **0.746** ± 0.432 | 7.505 ± 10.088 | **7.265** ± 8.939 |

Table 1: Quantitative segmentation results on the DRIVE database (2D) and on the IRCAD database (3D).

In these experiments, we have chosen the STARE dataset [66] consisting of 20 retinophotographies, which is similar to the target DRIVE dataset. In scenario 1, called MT STARE, we trained the MT using both the STARE dataset (with annotations) and the DRIVE dataset (without annotation). In scenario 2, called MT synth., we extracted the image backgrounds from the STARE dataset [66] and added synthetic curvilinear structures generated by the CCO algorithm, similarly to what is proposed in Lin *et al.* [13]. An example of such generated images is shown on Fig 4. Then we trained the MT using both the generated image (with annotations) and the DRIVE dataset (without annotation).

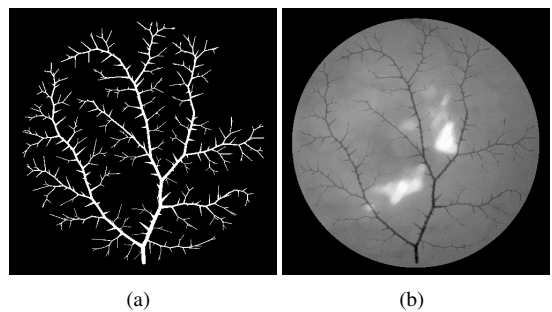

(a)  (b)

Figure 4: Simulation of a retinophotography (b) using a binary curvilinear structure (a) generated by the OpenCCO algorithm and an image background extracted from the STARE dataset [66].

| | Regularization | $\lambda$ |
|---|---|---|
| | TV | $0.0096 \pm 0.0072$ |
| | Directional TV | $0.0176 \pm 0.0145$ |
| 2D | $E_{\text{remove}}$ | $0.0072 \pm 0.0069$ |
| | $E_{\text{fill}}$ | $0.0114 \pm 0.0078$ |
| | $E_{\text{reco}}$ (with Diceloss) | $0.0085 \pm 0.0080$ |
| | $E_{\text{reco}}$ | $0.0089 \pm 0.0075$ |
| | TV | $0.0095 \pm 0.0112$ |
| 3D | Directional TV | $0.0119 \pm 0.0147$ |
| | $E_{\text{reco}}$ | $0.0084 \pm 0.0098$ |

Table 2: Mean ± standard deviation value of the regularization coefficient, $\lambda$, that were optimized for each approach.

### 4.1.3. Evaluation metrics

To evaluate the methods quantitatively, we computed metrics derived from the confusion matrix (True Positive TP, True Negative TN, False Positive FP, and False Negative FN). As vessel segmentation is an unbalanced problem, we computed: the sensitivity TPR, the precision PPV, the Dice and the Matthew Correlation Coefficient MCC (see Eqs. 12). We will refer to these metrics as volumetric metrics. They were computed inside the field of view of the retinal images or inside the liver mask of the CT-scans.

$$
\begin{aligned}
\text{TPR} &= \frac{\text{TP}}{\text{TP} + \text{FN}}, \\
\text{PPV} &= \frac{\text{TP}}{\text{TP} + \text{FP}}, \\
\text{Dice} &= \frac{2\text{TP}}{2\text{TP} + \text{FN} + \text{FP}}, \\
\text{MCC} &= \frac{\text{TP.TN} - \text{FP.FN}}{\sqrt{(\text{TP} + \text{FP})(\text{TP} + \text{FN})(\text{TN} + \text{FP})(\text{TN} + \text{FN})}}.
\end{aligned} \tag{12}
$$

None of these metrics give insight on the geometry or the connectivity of the curvilinear structures. Indeed, the difference between a connected and a disconnected result is usually only a few pixels, which does not change significantly volumetric metrics, but drastically reduce its interest for follow-up applications. To achieve an evaluation centered on the geometry of the curvilinear structures, we computed the ClDice [7], the 95$^{\text{th}}$ percentile of the Hausdorff Distance 95HD, and the Average Symmetric Surface Distance ASSD. We will refer to these metrics as geometric metrics. To evaluate the connectivity of our results, we introduced classic topological metrics called Betti's numbers: $\beta_0$, $\beta_1$ and

$\beta_2$. $\beta_0$ corresponds to the number of connected components, $\beta_1$ represents the number of tunnels, and $\beta_2$ denotes the number of cavities in a binary shape. We also added the Euler number, $\chi$, defined as a linear combination of the Betti's numbers:

$$
\chi = \beta_0 - \beta_1 + \beta_2. \tag{13}
$$

To evaluate the variation of these metrics in relation to their value in the annotated images, we computed the error ratio of $\beta_0$, $\beta_1$ and $\chi$ (see Eq. 14). We did not calculate the error ratio of $\beta_2$, as its value is 0 in the ground truth and in practice we did not observe cavities in the 3D segmentation results.

$$
\epsilon_{\text{M}} = \left| \frac{M - M_{\text{gt}}}{M_{\text{gt}}} \right|, \tag{14}
$$

with $M$ a topological metric (either $\beta_0$, $\beta_1$, or $\chi$) computed on a segmentation and $M_{\text{gt}}$ the same metric computed on its associated annotation. As these topological metrics are very sensitive to noise, we performed a post-processing before computing them. It consists in removing the small connected components (size less than 20 pixels in 2D and less than 90 pixels in 3D) and filling the small holes (size less than 10 pixels in 2D). This post-processing was neither applied to the qualitative results presented in this article nor before computing the volumetric and geometric metrics.

### 4.2. Results

In this section, we compare the results of our framework with the other strategies presented in section 4.1.2. Quantitative results are presented in Table 1, while some qualitative results are presented in Fig. 2 and in Fig. 3.

Qualitatively, the directional TV preserves more curvilinear fragments than TV, however these fragments are not connected, as shown by the similar values of $\epsilon_{\beta_0}$. Our segmentation results are smoother and demonstrate better connectivity compared to the TV and directional TV approaches. Quantitatively, our approach only slightly increases the volumetric metrics. However, this small increase does not adequately represent the importance of the improvement of the results. Indeed, it comes from the addition of a few pixels to reconnect the vessels, which is crucial for ensuring better global coherence and connectivity of the vascular networks both in 2D and 3D. This improvement is visually clear, and is demonstrated by the significant decrease in topological metrics. In particular, the error on the number of connected components of our results has decreased by almost 90% in 2D and 70% in 3D. Despite the capability of our reconnecting term to remove non-curvilinear fragments, we observe a few false reconnections in the results (see pink arrows in Fig 2 and in Fig 3). This may happen when artefacts are aligned or in the same orientation as a curvilinear structure.

| Method | Volumetric Metrics | | | | Geometric Metrics | | | Topological Metrics | | |
|---|---|---|---|---|---|---|---|---|---|---|
| | TPR | PPV | MCC | Dice | ClDice | 95HD | ASSD | $\epsilon_{\beta_0}$ | $\epsilon_{\beta_1}$ | $\epsilon_{\chi}$ |
| Before $G_{\text{reco}}$ | 0.967 ± 0.006 | 0.982 ± 0.005 | 0.971 ± 0.004 | 0.974 ± 0.004 | 0.959 ± 0.006 | 0.350 ± 0.477 | 0.221 ± 0.072 | 107.367 ± 71.883 | 0.113 ± 0.103 | 4.410 ± 1.746 |
| After $G_{\text{reco}}$ | 0.989 ± 0.003 | 0.977 ± 0.004 | 0.980 ± 0.003 | 0.983 ± 0.003 | 0.980 ± 0.005 | 0.375 ± 0.484 | 0.082 ± 0.014 | 17.301 ± 12.690 | 0.151 ± 0.135 | 0.543 ± 0.234 |
| $E_{\text{remove}}$ | 0.698 ± 0.054 | **0.825** ± 0.068 | 0.726 ± 0.039 | 0.753 ± 0.035 | 0.737 ± 0.045 | 16.044 ± 5.431 | **2.524** ± 0.915 | 20.758 ± 12.537 | 0.722 ± 0.122 | 1.937 ± 0.438 |
| $E_{\text{fill}}$ | 0.720 ± 0.050 | 0.793 ± 0.077 | 0.720 ± 0.047 | 0.751 ± 0.041 | 0.732 ± 0.048 | **15.428** ± 5.605 | 2.539 ± 1.041 | 4.208 ± 4.476 | 0.309 ± 0.173 | 0.474 ± 0.271 |
| $E_{\text{reco}}$ (w/o $\mathcal{D}_w$) | 0.716 ± 0.051 | 0.810 ± 0.071 | 0.728 ± 0.044 | 0.757 ± 0.038 | 0.745 ± 0.048 | 16.351 ± 5.852 | 2.583 ± 1.019 | 2.979 ± 2.523 | 0.371 ± 0.195 | 0.531 ± 0.239 |
| $E_{\text{reco}}$ | **0.725** ± 0.049 | 0.803 ± 0.069 | **0.729** ± 0.041 | **0.759** ± 0.036 | **0.744** ± 0.045 | 16.095 ± 5.235 | 2.550 ± 0.924 | **2.685** ± 2.767 | **0.316** ± 0.187 | **0.417** ± 0.232 |

Table 3: Ablation study on the DRIVE database (2D). The first two rows of the table present quantitative results for the evaluation of $G_{\text{reco}}$ alone applied to the 40 binary annotations of the DRIVE dataset, that were initially disconnected by our algorithm. The subsequent rows present the results of the ablation study of the segmentation framework with different versions of the regularization term, applied to the 40 images of the DRIVE dataset.

Moreover, our results present a slightly higher value for the distance-based metrics (95HD, ASSD) which may be due to the removal of small fragments that belong to vessels, but were detected as artefacts by our approach. However, the global vascular geometry is still better preserved as shown by the increase in the ClDice value.

The classic vesselness filters (RORPO and Frangi) yield segmentation volumes similar to the ones from the variational approaches, as shown by the close volumetric metrics. However, they exhibit much more artefacts, which increases the topological errors. Unsupervised deep learning methods achieved either an incomplete (MT STARE) or noisy (MT synth.) segmentation. However, the global architecture of the vascular network is relatively well preserved, as shown by the topological metrics. It is important to note that deep learning approaches show more promising results in a semi-supervised context where a few annotations are still available [67], which is out of the scope of this article.

## 4.3. Ablation study

To better understand the role of our reconnecting term $G_{\text{reco}}$, we conducted several experiments. First, we validated its behavior on disconnected binary curvilinear structures, independently of the segmentation variational scheme. Then, we analysed separately the design choices of this term. Finally, we analysed the influence of injecting earlier our term into the variational scheme. The qualitative results of the ablation study are presented in Fig 2(i-k) and in Fig 6. The quantitative evaluation is presented in Table 3.

### 4.3.1. Validation of $G_{reco}$

To validate our reconnecting term, we applied it on the DRIVE manual annotations that we had disconnected with our algorithm. In Fig. 6, most disconnections were reconnected and artefacts removed, as confirmed by the quantitative results presented in the first two lines of Table 3. All metrics showed significantly improvement after applying our reconnecting term, with particular emphasis on $\epsilon_{\beta_0}$, highlighting a good reconnecting behavior as expected.

### 4.3.2. $E_{fill}$ v.s. $E_{remove}$ v.s. $E_{reco}$

Our reconnecting term performs two tasks. Firstly, it fills small disconnections between curvilinear fragments that are close and in the same orientation. We refer to this task as *fill*. Secondly, it removes small components that are just noise and are not supposed to be reconnected to the main structure. We refer to this task as *remove*. In this section, we study these two tasks separately, and together to show the interest of their synergy.

We trained three different regularization terms: $G_{\text{fill}}$ which only fills disconnections between curvilinear fragments, $G_{\text{remove}}$ which only removes small fragments that are not aligned with curvilinear structures, and $G_{\text{reco}}$ which performs both tasks. These three models were trained with the same architecture presented in Section 3.3. The only difference is the ground truth that was used to train the models, which we generated according to the task of interest.

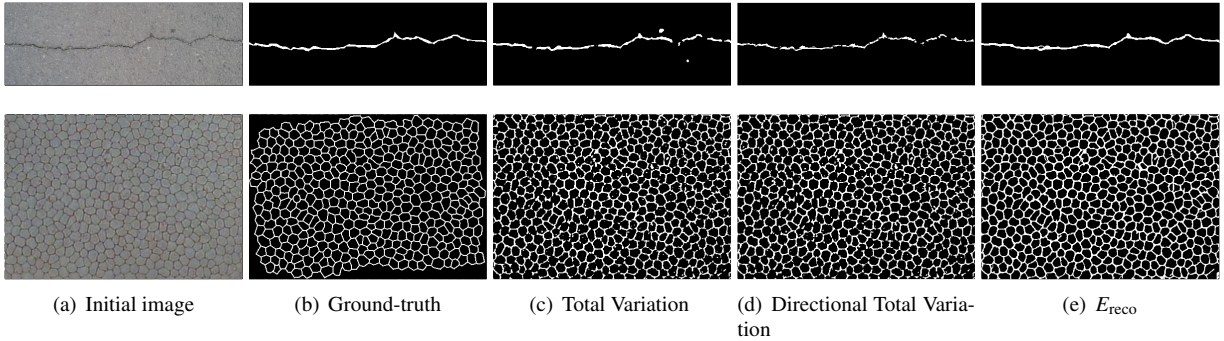

| (a) Initial image | (b) Ground-truth | (c) Total Variation | (d) Directional Total Variation | (e) $E_{\text{reco}}$ |

Figure 5: Segmentation of various curvilinear structures. The first row depicts an image from the road crack dataset [68, 69], and the second row shows an image from the porcine corneal cells dataset [70].

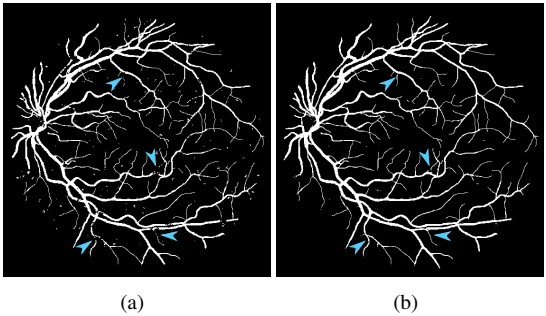

| (a) | (b) |

Figure 6: Test of our reconnecting term on the manual annotations of the Drive dataset.(a) the disconnected input image, (b) the result after applying $G_{\text{reco}}$ on (a).

Then, each term is added in the variational segmentation scheme, replacing the proximity operator of the regularization term ($E_{\text{fill}}$, $E_{\text{remove}}$, and $E_{\text{reco}}$). The results are presented in Table 3.

The results of $E_{\text{remove}}$ are very similar, both qualitatively and quantitatively, to the results of the segmentation with TV only. Removing non-curvilinear structure fragments is a type of denoising, a task already performed by TV, so it does not significantly improve the results on its own.

The results of $E_{\text{fill}}$ are similar based on the MCC to the results of TV and directional TV. However, a significant improvement is observed when looking at the topological metrics. In particular, the error on the ber of connected components ($\epsilon_{\beta_0}$) is almost divided by 5 compared with $E_{\text{remove}}$ or TV. This shows that the reconnection worked as intended.

The mean and standard deviation of the topological errors of $E_{\text{reco}}$ is improved compared to $E_{\text{fill}}$ alone, and in particular $\epsilon_{\beta_0}$ is significantly improved. Moreover, we observed that, by including a denoising targeting non-curvilinear structure fragments, the strength of the non-specific denoising TV can be reduced (see Table 2), allowing for more fragments to potentially be reconnected later by the reconnecting task.

### 4.3.3. Weighted Dice loss

As described in Eq. 2, we use a combination of two Dice losses to train our neural network. To prove the interest of this loss, we have trained 2D reconnecting terms both with and without the weighted Dice loss and plugged them into our segmentation framework. Quantitatively, we can observe that including the combination of Dice losses slightly improved the segmentation results, in particular regarding the topological metrics. While the difference is subtle in 2D, the 3D results show a much greater improvement with the addition of this extra loss.

### 4.3.4. Influence of $\alpha$

As described in the section 3.4.2, our reconnecting term must be injected when its input is almost binary. In the previous experiments, we set $\alpha = 500$.

In this section, we investigate the influence of the parameter $\alpha$ on the results of our approach. We systematically varied $\alpha$ across the range of $[0, 500]$ with increments of 50. The results of this experiment are presented in Fig. 4.3.4.

---

[2]https://github.com/cuilimeng/CrackForest-dataset

When the reconnecting model is introduced at the beginning of the optimization scheme ($\alpha = 0$), the reconnecting term is applied to an image that deviates significantly from the distribution it has been trained on, *i.e.*, far from a binary image. Consequently, it produces increasingly aberrant results, leading to poor segmentations.

Starting from $\alpha = 50$, the MCC and the ASSD stabilize, which coincides with the binarization of the image. Although the MCC and ASSD exhibit minimal variation beyond $\alpha = 50$, $\epsilon_{\beta_0}$ continues to decrease slowly. By $\alpha = 200$, the results reach a plateau. This observation suggests that the optimal $\alpha$ lies around 200. Moreover, it is noteworthy that increasing $\alpha$ beyond this value does not lead to degradation in results, underscoring the robustness of our approach with respect to this parameter.

### 4.4. Generalization on various curvilinear structures

By training our regularization operator on synthetic datasets, and successfully applying it to real vascular images, we have demonstrated that our learned reconnecting term is independent of the target dataset. In this section, we further illustrate this robust generalization behavior by applying our approach to non-vascular structures. Specifically, we evaluate our 2D reconnecting term, trained on the OpenCCO dataset, on two additional datasets: one comprising 118 images of road cracks [68, 69], and another containing 30 images of cells from porcine corneal endothelium [70]. We compare the performance of our term with that of the Total Variation and the directional Total Variation. Results are presented in Table 4 and Figure 5.

Despite being trained on synthetic binary vascular trees, our approach consistently produces high-quality results, notably enhancing the connectivity of the segmentation. This shows the effectiveness of our learned reconnecting term, which, by operating on binary structures and prioritizing the geometry of curvilinear structures, exhibits strong generalizability across diverse applications.

It is important to note that our approach rely on the Chan *et al.* segmentation data fidelity term [60] which is based on the assumption that the background of the input image is homogeneous. In our experiments, we consistently preprocessed the datasets to ensure this property. However, in some applications, this preprocessing is not enough to homogenize the image background. This is in particular the case when the image present multiple non-curvilinear objects with a high contrast. In this case, our framework can not be applied directly, and additional preprocessing should be considered such as vesselness filters [42].

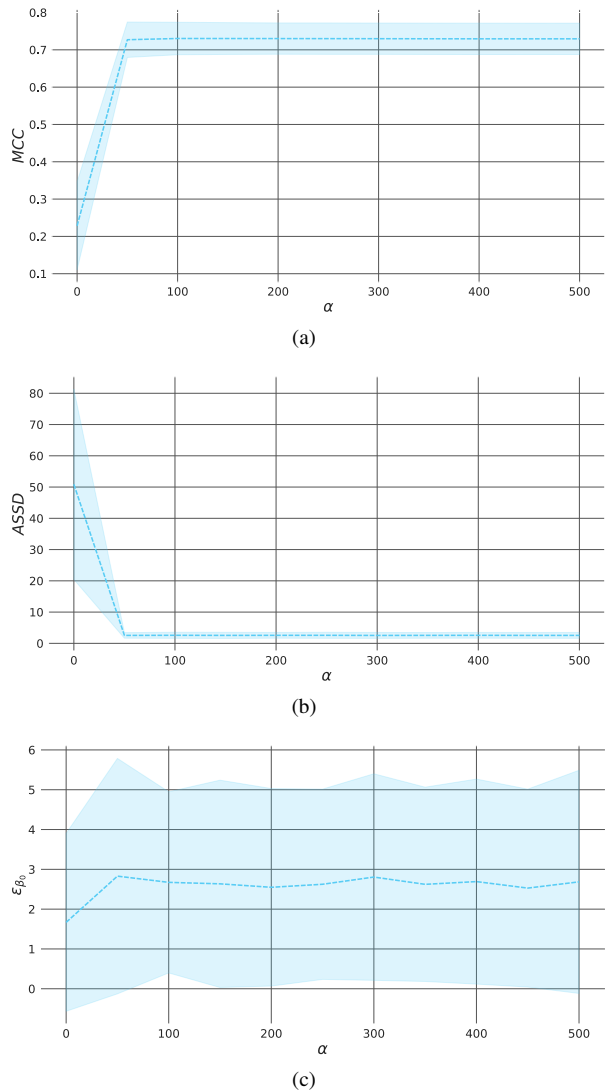

Figure 7: Variation of MCC (a), ASSD (b), and $\epsilon_{\beta_0}$ (c) on the DRIVE dataset in function of the iteration $\alpha$ from which the $G_{\text{reco}}$ model is injected into the optimization scheme.

| | Method | MCC | ASSD | $\epsilon_{\beta_0}$ |
|---|---|---|---|---|
| **Cracks** | TV | 0.517 | 14.358 | 15.928 |
| | | ± 0.171 | ± 12.322 | ± 26.183 |
| | Directional TV | 0.557 | **9.412** | 10.043 |
| | | ± 0.141 | ± 9.426 | ± 10.816 |
| | $E_{reco}$ | **0.605** | 9.691 | **2.161** |
| | | ± 0.153 | ± 11.047 | ± 3.296 |
| **Cells** | TV | 0.604 | 1.280 | 45.033 |
| | | 0.038 | ± 0.271 | ± 33.291 |
| | Directional TV | 0.606 | 1.273 | 46.700 |
| | | ± 0.038 | ± 0.276 | ± 35.109 |
| | $E_{reco}$ | **0.637** | **1.261** | **0.267** |
| | | ± 0.030 | ± 0.268 | ± 0.772 |

Table 4: Quantitative segmentation results on the cracks and porcine corneal cells.

## 5. Conclusion

In this article, we proposed a comprehensive plug-and-play framework for curvilinear structure segmentation that includes a dedicated constraint for the preservation of connectivity. It includes an algorithm for generating realistic pairs of connected/disconnected curvilinear structures and a reconnecting regularization operator that can be learned from a synthetic dataset. Contrary to classic deep learning segmentation approaches, our approach does not need annotations of the dataset of interest. Finally, we proposed a strategy to include this term into a variational segmentation scheme.

In our study, we demonstrated the efficacy of our regularization term through an ablation analysis. By learning a regularization term that can identify and remove non-curvilinear structure fragments, as well as reconnect true curvilinear structure fragments, significant improvements can be achieved in terms of global structure connectivity. This is evidenced by the topological metrics both in 2D and 3D.

In the future, it would be interesting to provide guarantees on the convergence of our plug-and-play approach. In particular, by learning our reconnecting term to be a maximally monotone operator [71]. This would enhance the robustness and reliability of our approach. Furthermore, we see potential in including our reconnecting term as a fixed layer of an end-to-end segmentation network to enforce connectivity in a supervised setting.

## CRediT authorship contribution statement

**Sophie Carneiro-Esteves:** Conceptualization, Methodology, Software, Validation, Formal analysis, Investigation, Data Curation, Writing - Original Draft, Visualization. **Antoine Vacavant:** Conceptualization, Writing - Review & Editing, Supervision, Project administration, Funding acquisition. **Odyssée Merveille:** Conceptualization, Writing - Review & Editing, Supervision, Project administration, Funding acquisition.

## Acknowledgment

This work was supported by Agence Nationale de la Recherche (ANR-22-CE45-0018, ANR-18-CE45-0018), LABEX PRIMES (ANR-11-LABX-0063). It granted access to the HPC resources of IDRIS under the allocations 2022-AD011013887 made by GENCI.